\documentclass{article}
\usepackage{amssymb}
\usepackage{amsmath}
\usepackage[dvips]{graphicx}
\usepackage{fleqn,ams}
\usepackage{verbatim}
\usepackage[cp1251]{inputenc}
\usepackage[dvips]{graphicx}
\begin{document}
\begin{center}
{\bf\Large Dinamic model of spherical perturbations in Friedmann Universe} \\[12pt]
Yu.G. Ignatyev, N. Elmakhi\\
Tatar State Humanitarity Pedagocical University\\ 1 Mezhlauk St.,
Kazan 420021, Russia
\end{center}

\begin{abstract}
A selfconsistent system of equations describing evolution of
linear spherically symmetric perturbations in Fridmann world by an
arbitrary equation of state is obtained. The perturbations
singular part corresponding to massive particle-like source is
extracted, an evolutionary equation for mass of this source is
obtained and exactly solved. An exact solution of evolutionary
equations of state for perturbations by an arbitrary equation of
state is constructed.
\end{abstract}

\section{Introduction}

In a number of papers of one of the Authors collaboratively with
A.A. Popov Ref.[1]-[3] a theory of spherical perturbations of the
Friedmann world in con\-nec\-tion with the necessity of developing
the relativistic kinetic theory considering gravi\-ta\-tio\-nal
interactions was constructed. The procedure applied for the
kinetic equations derivation of averaging local fluctuations of
the gravitational field pointed out by one of the Authors an
interesting fact: the average quadrature fluctuations of the
gravitational field play a role of  tensor of energy-momentum of
an ideal fluid  with extremely-hard equation of state [3]. In the
paper of one of the Authors collaboratively with A.A. Popov exact
solutions of the equations for the spherically symmetric
perturbations of the ultrarelativistic Friedmann world with an
arbitrary curve index [4] were obtained. However, the analysis of
the obtained solutions and the calculation of the average
quadrature corrections to the Friedmann metrics were not
fulfilled. In connection with the problem of dark energy and dark
materia in cosmology the task of possible alteration of
macroscopic equation of state of the Friedmann universe by the
local gravitational interactions becomes actual, as far as the
potential energy on the gravitational interactions is negative, it
can correspond to a negative macroscopic pressure. In the present
work we start investigating this possibility  by fulfilling
subsequent maintaining a dynamic theory of the Friedmann universe
subject to the local gravitational perturbations.

\section{Spherically symmetric space-time }
\subsection{ Space symmetry and energy-momentum tensor }

Let us study the space-time with spherical symmetry which metrics
in the isotropic frame of reference\footnote{That is in the frame
of reference in which the metrics of   three-dimention space takes
a conform plane form } $(r,\theta ,\varphi ,\eta )$, where $\eta $
is the time coordinate  and $r$ is a derivative one, can be
written in the form

\begin{equation} \label{GrindEQ__1_}
ds^{2} =e^{\nu } d\eta ^{2} -e^{\lambda }
[dr^{2} +r^{2} (d\theta ^{2} +\sin ^{2} \theta d\varphi ^{2} )],
\end{equation}
where
\[\lambda =\lambda (r,\eta );\quad \nu =\nu (r,\eta )\]
are arbitrary scalar functions of their arguments. Isotropic
coordinates are con\-ve\-ni\-ent because within these coordinates
the three-dimension space metrics takes an explicitly conform
plane form. As it is known the metrics \eqref{GrindEQ__1_} admits
a group of  $G_{3} $ rotations with the Killing's vectors (see,
for example Ref. [7])
\begin{equation} \label{GrindEQ__2_}
\begin{array}{l} {\mathop{\mathop{\xi }\limits_{1} \, }\nolimits^{i} =
(0,\sin \varphi ,\theta \cos \varphi ,0),{\rm \; \; }
\mathop{\mathop{\xi }\limits_{2} \, }\nolimits^{i} =(0,-\cos \varphi ,\theta \sin \varphi ,0),} \\
{\mathop{\mathop{\xi }\limits_{3} \, }\nolimits^{i} =(0,0,1,0),} \end{array}
\end{equation}
Therefore in consequence of the Einstein equations the space-time
symmetry is taken after by the energy-impulse tensor
\begin{equation} \label{GrindEQ__3_}
\mathop{L}\limits_{\mathop{\xi }\limits_{\alpha } \, } \, T^{ik} =0,
\quad (\alpha =\overline{1,3}),
\end{equation}
where $\mathop{L}\limits_{\xi } \, $ is the Lie's derivative in
the direction of $\xi $ (see, for example Ref. [7])
\begin{equation} \label{GrindEQ__4_}
\mathop{L}\limits_{\xi } \, A^{i} =A_{,k}^{i} \xi ^{k} +A^{k} \xi _{,k}^{i}
\end{equation}
the other tensor derivatives are processed in the same way as the
Lie's derivative. Thus under the conditions of spherical symmetry
of space-time the energy-impulse tensor, as it is known, takes the
form of the energy-impulse tensor of an ideal isotropic fluid
\begin{equation} \label{GrindEQ__5_}
T^{ik} =(\varepsilon +p)u^{i} u^{k} -pg^{ik} ,
\end{equation}
Where the scalars $\varepsilon (r,\eta )$ and  $p(r,\eta )$ are an
energy density and a fluid pressure respectively and $u^{i}$ is a
singular time-like vector of this fluid dynamic velocity
\begin{equation} \label{GrindEQ__6_}
g_{ik} u^{i} u^{k} =1,
\end{equation}
whereas
\begin{equation} \label{GrindEQ__7_}
u^{i} =(u^{r} (r,\eta ),0,0,u^{\eta } (r,\eta )).
\end{equation}
Assuming
\[u^{r} =vu^{\eta } e^{{\tfrac{\nu -\lambda }{2}} } ,\quad v^{2} <1,\]
where $v(r,\eta )$ is a radial three-dimension velocity of fluid,
we get from \eqref{GrindEQ__6_} and \eqref{GrindEQ__5_}
\begin{equation} \label{GrindEQ__8_}
u^{\eta } =e^{-{\tfrac{\nu }{2}} } \frac{1}{\sqrt{1-v^{2} } } ;
\end{equation}
\[T_{4}^{1} =(\varepsilon +p)e^{(\nu -\lambda )/2} \frac{v}{1-v^{2} } ;
\quad T_{4}^{4} =\frac{\varepsilon +v^{2} p}{1-v^{2} } ;\]

\begin{equation} \label{GrindEQ__9_}
T_{1}^{1} =-\frac{\varepsilon v^{2} +p}{1-v^{2} } ;\quad T_{2}^{2} =T_{3}^{3} =-p.
\end{equation}
Thus the following algebraic relations are valid
\begin{equation} \label{GrindEQ__10_}
\begin{array}{l} {T_{1}^{1} +T_{4}^{4} =\varepsilon -p;} \\ {} \\ {T_{4}^{4} -T_{1}^{1} =
(\varepsilon +p)\frac{1+v^{2} }{1-v^{2} } ;} \\ {} \\ {T=T_{1}^{1} +T^{2} +T_{3}^{3} +T_{4}^{4} =
\varepsilon -3p.} \end{array}
\end{equation}

\subsection{Einstein Equations }
Nontrivial Einstein equations relatively the metrics
\eqref{GrindEQ__1_} have the form (see, for example Ref.
[6])\footnote{In order to obtain these equations in the formulae
of Ref. [6] it is necessary to assume  $\mu =\lambda +2\ln r$ .}:
\begin{eqnarray} \label{GrindEQ__11_}
\frac{1}{2} e^{-\lambda } \left(\frac{\mathop{\lambda
'}\nolimits^{2} }{2} + \lambda '\nu '+\frac{2}{r} (\lambda '+\nu
')\right)-e^{-\nu } \left(\ddot{\lambda }-\frac{1}{2} \dot{\lambda
}\dot{\nu }+\frac{3}{4} \mathop{\dot{\lambda }}\nolimits^{2}
\right)\nonumber\\=8\pi \frac{\varepsilon v^{2} +p}{1-v^{2} }
\quad (=-8\pi T_{1}^{1} );
\end{eqnarray}
\begin{eqnarray} \label{GrindEQ__12_}
\frac{1}{4} e^{-\lambda } \left[2(\lambda ''+\nu '')+\mathop{\nu
'}\nolimits^{2} +\frac{2}{r} (\lambda '+\nu ') \right]-e^{-\nu }
\left(\ddot{\lambda }-\frac{1}{2} \dot{\lambda }\dot{\nu
}+\frac{3}{4} \mathop{\dot{\lambda }}\nolimits^{2}
\right)\nonumber\\=8\pi p\quad (=-8\pi T_{2}^{2} ),
\end{eqnarray}
\begin{equation} \label{GrindEQ__13_}
-e^{-\lambda } \left(\lambda ''+\frac{1}{4} \mathop{\lambda
'}\nolimits^{2} + \frac{2}{r} \lambda '\right)+\frac{3}{4} e^{-\nu
} \mathop{\dot{\lambda }}\nolimits^{2} = 8\pi \frac{\varepsilon
+v^{2} p}{1-v^{2} } \quad (=8\pi T_{4}^{4} );
\end{equation}
\begin{equation} \label{GrindEQ__14_}
\frac{1}{2} e^{-\lambda } (2\dot{\lambda }'-\nu '\dot{\lambda
})=8\pi (\varepsilon +p)e^{(\nu -\lambda )/2} \frac{v}{1-v^{2} }
\quad(=8\pi T_{4}^{1} ),
\end{equation}
where $f'$ is a derivative from the function $f$ by the radial
variable $r$ and $\dot{f}$ is a derivative by the time variable
$\eta $ and the universal set of units is assumed everywhere in
which $G=c=\hbar =1$. Subtracting from the both parts of Eq.
\eqref{GrindEQ__11_} the corresponding parts of  Eq.
\eqref{GrindEQ__12_} subject to \eqref{GrindEQ__9_} we get the
consequence
\begin{eqnarray} \label{GrindEQ__15_}
\frac{1}{2} e^{-\lambda } \left[\frac{1}{2} \mathop{\lambda
'}\nolimits^{2} + \lambda '\nu '-\frac{1}{2} \mathop{\nu
'}\nolimits^{2} + \frac{1}{r} (\lambda '+\nu ')-(\lambda ''+\nu
'')\right]\nonumber\\=8\pi (\varepsilon +p)\frac{v^{2} }{1-v^{2} }
.
\end{eqnarray}

\section{Background space-time }
Whithin the isotropic spherical coordinates the nonperturbed
gravitational field is described by the metrics of homogeneous
isotropic universe
\begin{equation} \label{GrindEQ__16_}
ds^{2} =a^{2} (\eta )\left(d\eta ^{2} -\frac{1}{\rho ^{2} (r)} [dr^{2} +r^{2}
(d\theta ^{2} +\sin ^{2} \theta d\varphi ^{2} )]\right),
\end{equation}
where
\begin{equation} \label{GrindEQ__17_}
\rho (r)=1+\frac{1}{4} kr^{2} ,
\end{equation}
And the curvature index  $k=0$ is for the space-plane universe,
$k=\pm 1$ is for the universe with positive and negative curvature
of thee-dimensional space respectively. Whereas $r$ and $\eta $
are dimensionless  variables and the scale factor $a(\eta )$ has
length dimension.

Thus in a nonperturbed state the introduced scalar metrics
functions $\lambda $ and $\nu $ are equal to
\begin{equation} \label{GrindEQ__18_}
\nu _{0} =\ln a^{2} (\eta );\quad \lambda _{0} =\ln \mathop{\left(\frac{a(\eta )}{\rho (r)}
\right)}\nolimits^{2} ,
\end{equation}
Whereas in consequence of time space homogeneity
\begin{equation} \label{GrindEQ__19_}
p=p_{0} (\eta );\quad \varepsilon =\varepsilon _{0} (\eta );\quad v_{0} =0.
\end{equation}
Substituting Eqs. \eqref{GrindEQ__18_} and \eqref{GrindEQ__19_}
into the Einstein Eqs. \eqref{GrindEQ__11_}-\eqref{GrindEQ__15_}
we obtain a set of equations describing the Friedmann universe
dynamics
\begin{equation} \label{GrindEQ__20_}
\frac{1}{a^{2} } \left(\frac{\mathop{\dot{a}}\nolimits^{2} }{a^{2} } +k\right)=\frac{8\pi }{3} \varepsilon ;
\end{equation}
\begin{equation} \label{GrindEQ__21_}
\frac{1}{a^{2} } \left(2\frac{\ddot{a}}{a} -\frac{\mathop{\dot{a}}\nolimits^{2} }{a^{2} } +k\right)=-8\pi p.
\end{equation}
As it is known the second of these equations \eqref{GrindEQ__21_}
can be substituted by the algebraic-differential consequence of 1
and 2 (see, for example Ref. [6])
\begin{equation} \label{GrindEQ__22_}
\dot{\varepsilon }+3\frac{\dot{a}}{a} (\varepsilon +p)=0.
\end{equation}
Here it is convenient to pass on from the time variable $\eta $ to
the physical time $t$ by the formula
\begin{equation} \label{GrindEQ__23_}
a(\eta )d\eta =dt,\quad \Rightarrow t=\smallint a(\eta )d\eta ;
\end{equation}
whereas
\begin{equation} \label{GrindEQ__24_}
\frac{\partial }{\partial \eta } =a\frac{\partial }{\partial t} \to \; \dot{f}=a\mathop{\dot{f}}\nolimits_{t} .
\end{equation}
Then the independent Einstein equations can be written in a more
compact form (see, for example Ref. [6])
\begin{equation} \label{GrindEQ__25_}
\frac{1}{a^{2} } (\mathop{\dot{a}}\nolimits_{t}^{2} +k)=\frac{8\pi }{3} \varepsilon ;
\end{equation}
\begin{equation} \label{GrindEQ__26_}
\mathop{\dot{\varepsilon }}\nolimits_{t} +3\frac{\mathop{\dot{a}}\nolimits_{t} }{a} (\varepsilon +p)=0,
\end{equation}
If we know the equation of state, that is the relation of the form
\begin{equation} \label{GrindEQ__27_}
p=p(\varepsilon ),
\end{equation}
then  Eq. \eqref{GrindEQ__22_} is integrated in quadratures
\begin{equation} \label{GrindEQ__28_}
-3\ln a=\smallint \frac{d\varepsilon }{\varepsilon +p(\varepsilon )} +{\rm Const}.
\end{equation}
Substituting the solution \eqref{GrindEQ__28_} into  Eq.
\eqref{GrindEQ__20_} we obtain a closed differential equation of
order 1 relatively $\varepsilon (\eta )$. In the case of
barotropic equation of state\footnote{We call the reader's
attention to the difference ,   $k$  is a curvature index  and
$\kappa $  (kappa) is a bartrop coefficient  .}
\begin{equation} \label{GrindEQ__29_}
p=\kappa \varepsilon
\end{equation}
Eq. \eqref{GrindEQ__24_} is easily integrated
\begin{equation} \label{GrindEQ__30_}
\varepsilon =c_{1} a^{-3(\kappa +1)} ,
\end{equation}
and  Eq. \eqref{GrindEQ__20_} is integrated in  quadratures
\begin{equation} \label{GrindEQ__31_}
\int \frac{da}{a\sqrt{\frac{8\pi c_{1} }{3} a^{2-3(\kappa +1)} -k} }  =c_{2} \eta ,
\end{equation}
where $c_{1} $ and $c_{2} $ are arbitrary constants. The pointed
out equations are integrated in the elementary functions for the
early Universe ($t\to 0$). As it is known in this case the
behavior of the solutions does not depend on the curvature $k$
(see, for example Ref. [6]) and does not differ from the behavior
of the  solutions for a space plane Universe  ($k=0$)
\begin{equation} \label{GrindEQ__32_}
a=a_{1} \eta ^{\frac{2}{3\kappa +1} } ;\; \varepsilon =c_{1} a_{1}^{-3(1+\kappa )}
\eta ^{-\frac{6(1+\kappa )}{3\kappa +1} } ,\quad \kappa +1\ne 0,
\end{equation}
Where the constants $a_{1} $  and  $c_{1} $ are connected by the relation

\begin{equation} \label{GrindEQ__33_}
a_{1} =\mathop{\left((3\kappa +1)\sqrt{\frac{2\pi c_{1} }{3} } \right)}\nolimits^{\frac{2}{3\kappa +1} } ,\quad 1+3\kappa \ne 0.
\end{equation}
In doing so let us reduce a relation between the variables $t$ and $\eta $ from Eq. \eqref{GrindEQ__23_}

\begin{equation} \label{GrindEQ__34_}
t=a_{1} \frac{3\kappa +1}{3(1+\kappa )} \eta ^{\frac{3(1+\kappa )}{3\kappa +1} } .
\end{equation}
Subject to \eqref{GrindEQ__33_} and \eqref{GrindEQ__34_}  let us reduce from \eqref{GrindEQ__32_}

\begin{equation} \label{GrindEQ__35_}
\varepsilon =\frac{1}{2\pi (\kappa +1)^{2} t^{2} }
\end{equation}
Note the solution of the Einstein equations by $\kappa =-1/3$ is specific

\[a=c_{2} e^{\sqrt{\frac{8\pi }{3} c_{1} } \eta } ,\quad (\kappa =-\frac{1}{3} ,\; \eta \in (-\infty ,+\infty ).\]
However this property is a coordinate only. Really, by passing on from the time variable  $\eta $ to the physical time $t$

\[t=\frac{c_{2} }{\sqrt{{\tfrac{8\pi }{3}} c_{1} } } \; e^{\sqrt{\frac{8\pi }{3} c_{1} } \eta } \]
we obtain  $a\sim t$ and the energy dense formula \eqref{GrindEQ__35_} in which it is necessary to substitute $\kappa =-1/3$ only.  By $\kappa =-1$ we get from \eqref{GrindEQ__32_} the so-called inflational solution

\[a=-\frac{1}{\Lambda \eta } ,\quad t=-\ln \eta ;\]

\begin{equation} \label{GrindEQ__36_}
a=a_{1} e^{\Lambda t} ;\quad \varepsilon =\frac{3\Lambda ^{2} }{8\pi } ={\rm const}.
\end{equation}

The solutions \eqref{GrindEQ__33_} corresponding to the values
$\kappa <-1$ describe the so-called dark materia.

\section{Linear spherically-symmetric perturbations of Friedmann
space-time}

\subsection{Equations for spherically-symmetric perturbations}
Let us consider now the small spherically-symmetric
perturbations of isotropic cosmological solution
\eqref{GrindEQ__18_} assuming

\begin{equation} \label{GrindEQ__37_}
\begin{array}{cc} {\lambda =\ln a^{2} (\eta )+\delta \lambda ;} & {\nu =\ln a^{2} (\eta )+\delta \nu ;} \\ {} & {} \\ {p=p_{0} (\eta )+\mathop{\left. \frac{dp}{d\varepsilon } \right|}\nolimits_{\varepsilon _{0} } \delta \varepsilon ;} & {\varepsilon =\varepsilon _{0} (\eta )+\delta \varepsilon ,} \end{array}
\end{equation}
Where the scalar functions $\delta \lambda (r,\eta )$, $\delta \nu (r,\eta )$, $\delta \varepsilon (r,\eta )$and $v(r,\eta )$ will be assumed small of order 1 by smallness. Substituting \eqref{GrindEQ__37_} into Eq. \eqref{GrindEQ__15_} in the first approximation by smallness of the perturbations  $\delta \lambda ,\delta \nu ,\delta \varepsilon $ and $v$ we obtain one closed equation relatively the function $\delta \lambda +\delta \nu $

\begin{equation} \label{GrindEQ__38_}
\frac{\partial }{\partial r} \left(\frac{\rho (r)}{r} (\lambda +\nu )'\right)=0.
\end{equation}
Integrating Eq. \eqref{GrindEQ__38_} we get (see also Ref.[4])

\begin{equation} \label{GrindEQ__39_}
\lambda +\nu =\left\{\begin{array}{cc} {C_{1} (\eta )+C_{2} (\eta )r^{2} ,} & {k=0;} \\ {} & {} \\ {C_{1} (\eta )+\frac{C_{2} (\eta )}{\rho (r)} ;} & {k=\pm 1,} \end{array}\right.
\end{equation}
where $C_{1} (\eta )$and $C_{2} (\eta )$ are arbitrary functions.

\noindent Further we shall search only the solutions of the Einstein equations of $C^{1} $ class, which out of some sphere coincide with homogeneous isotropic nonperturbed solution

\begin{equation} \label{GrindEQ__40_}
\begin{array}{l} {\mathop{\left. \lambda (r,\eta )\right|}\nolimits_{rr_{0} (\eta )} =\lambda _{0} (r,\eta );\quad \mathop{\left. \nu (r,\eta )\right|}\nolimits_{rr_{0} (\eta )} =\nu _{0} (\eta ),} \\ {\mathop{\left. \lambda '(r,\eta )\right|}\nolimits_{rr_{0} (\eta )} =\mathop{\lambda '}\nolimits_{0} (r,\eta );\quad \mathop{\left. \nu '(r,\eta )\right|}\nolimits_{rr_{0} (\eta )} =\mathop{\nu '}\nolimits_{0} (\eta ).} \end{array}
\end{equation}
Such solutions correspond to the retarded solutions of hyperbolic type equations. The physical sense of the solutions we shall discuss later. Then according to \eqref{GrindEQ__39_} it should take place (see Ref. [4])

\noindent

\begin{equation} \label{GrindEQ__41_}
\delta \lambda +\delta \nu =0,\Rightarrow \; \delta \lambda =-\delta \nu .
\end{equation}

\noindent In the paper  [4] spherically symmetric perturbations
are studied in the ultra\-re\-la\-ti\-vis\-tic universe ($\kappa
=1/3$) only, however in doing that solutions of linearized
Einstein equations for all types of Friedmann universe were
obtained. In the present paper we will restrict ourselves with the
case of  the space-plane universe ($k=0$), but the barotrop
coefficient $\kappa $ will be considered to be arbitrary. Thus,
subject to the background Einstein equations
\eqref{GrindEQ__20_}-\eqref{GrindEQ__21_}, which in the case of
$k=0$ have the consequence

\begin{equation} \label{GrindEQ__42_}
2\frac{\ddot{a}}{a} =\frac{\mathop{\dot{a}}\nolimits^{2} }{a^{2} } (1-3\kappa ),
\end{equation}
we get a closed set of three differential Einstein equations linearized around the background  solution \eqref{GrindEQ__16_},  relatively the three unknown variables $\delta \nu (r,\eta )$, $\delta \varepsilon (r,\eta )$and$v(r,\eta $)

\begin{equation} \label{GrindEQ__43_}
\delta \ddot{\nu }+3\delta \dot{\nu }\frac{\dot{a}}{a} -3\kappa \delta \nu \frac{\mathop{\dot{a}}\nolimits^{2} }{a^{2} } =8\pi a^{2} \delta p;
\end{equation}

\begin{equation} \label{GrindEQ__44_}
3\delta \dot{\nu }\frac{\dot{a}}{a} +3\delta \nu \frac{\mathop{\dot{a}}\nolimits^{2} }{a^{2} } -\frac{1}{r^{2} } \frac{\partial }{\partial r} r^{2} \frac{\partial }{\partial r} \delta \nu =-8\pi a^{2} \delta \varepsilon ;
\end{equation}

\begin{equation} \label{GrindEQ__45_}
\frac{1}{a^{3} } \frac{\partial }{\partial \eta } a\delta \nu '=-8\pi \varepsilon _{0} (1+\kappa )v.
\end{equation}
The latter of this set of equations \eqref{GrindEQ__45_} is a definition of the radial velocity $v(r,\eta )$. One of the equations \eqref{GrindEQ__43_} and \eqref{GrindEQ__44_} determines the density energy perturbation $\delta \varepsilon (r,\eta )$.

\subsection{Derivation of particlelike solutions}

Further assuming the investigation of  particle-like solutions of
perturbation equations also, let us study canonic equations of
motion of gravitating classical point particle in the
gravitational field to which the $\delta $-like energy density
cor\-res\-ponds. As a result of the two competitive processes --
the accretion  of material environment and the reverse process --
evaporation of substance the mass of the classic point particle in
the material environment cannot be constant. Therefore let us
write the Hamilton invariant function of massive particle in the
form\footnote{See the details in Ref. [8].}

\[H(x,P)=\sqrt{g^{ik} P_{i} P_{k} } -m,\quad (=0),\]
where $m=m(s)$ is a scalar function. From \eqref{GrindEQ__45_} we get the normalization

\noindent ratio

\begin{equation} \label{GrindEQ__46_}
(P,P)=m^{2} (s).
\end{equation}

\noindent The relativistic canonical equations of  particle motion take the form

\begin{equation} \label{GrindEQ__47_}
\frac{dx^{i} }{ds} =\frac{\partial H}{\partial P_{i} } ;\quad \frac{dP_{i} }{ds} =-\frac{\partial H}{\partial x^{i} } .
\end{equation}
From the first couple of the canonical equations subject to the normalization ratio we obtain

\begin{equation} \label{GrindEQ__48_}
\frac{dx^{i} }{ds} =\frac{P^{i} }{m} \Rightarrow g_{ik} \frac{du^{i} }{ds} \frac{du^{k} }{ds} =1.
\end{equation}
The second couple of the canonical equations of motion gives the Lagrangrian equations of classical massive particle of  variable mass

\begin{equation} \label{GrindEQ__49_}
\frac{d^{2} x^{i} }{ds^{2} } +\Gamma _{jk}^{i} \frac{dx^{j} }{ds} \frac{dx^{k} }{ds} =(\ln m)_{,k} \left(g^{ik} -\frac{du^{i} }{ds} \frac{du^{k} }{ds} \right).
\end{equation}

\noindent In spherically symmetric metrics the motion equations solution \eqref{GrindEQ__48_} which does not break the spherical symmetry, is a time line which corresponds to the particle rest state at the origin of the coordinates

\begin{equation} \label{GrindEQ__50_}
r=0,\; x^{4} =\eta ,
\end{equation}
in doing so the mass at rest can be an arbitrary function of the coordinate time

\begin{equation} \label{GrindEQ__51_}
m=m(\eta ).
\end{equation}
Let us write down the density energy corresponding to its singular part in invariant form

\begin{equation} \label{GrindEQ__52_}
\delta \varepsilon _{m} =m(\eta )\delta (r),
\end{equation}
where $\delta (r)$  is Dirac invariant $\delta $-function in spherical coordinates perceived in the sense of integral ratio

\begin{equation} \label{GrindEQ__53_}
\smallint d^{3} V\delta ^{3} (x)=a^{3} \smallint d\Omega \int _{0}^{r_{0} } \, r^{2} dr\delta (r)=4\pi a^{3} \int _{0}^{r_{0} } \, \delta (r)r^{2} dr=1,
\end{equation}
so that

\begin{equation} \label{GrindEQ__54_}
\smallint d^{3} V\delta \varepsilon _{m} =m(\eta ).
\end{equation}
Temporally abandoning the time derivatives by $\eta $ in the left part of  Eq. \eqref{GrindEQ__43_} we get then the following equation for the singular part corresponding to the singular part of density

\begin{equation} \label{GrindEQ__55_}
\frac{1}{r^{2} } \frac{\partial }{\partial r} r^{2} \frac{\partial }{\partial r} \delta \nu =8\pi a^{2} m(\eta )\delta (r).
\end{equation}
Multiplying the both parts \eqref{GrindEQ__55_} by $ar^{2} dr$ and integrating, then integrating by parts in the equation left part and at the same time assuming

\begin{equation} \label{GrindEQ__56_}
\mathop{\lim }\limits_{r\to 0} r^{2} \frac{\partial \delta \nu }{\partial r} =0
\end{equation}
we obtain

\begin{equation} \label{GrindEQ__57_}
ar^{2} \frac{\partial \delta \nu }{\partial r} =2m(\eta ).
\end{equation}
Integrating this equation one more time we get

\[\delta \nu =-\frac{2m(\eta )}{ar} .\]
Thus the ratio similar to the known one subject to redetermination of the invariant $\delta $-function \eqref{GrindEQ__53_} takes place

\begin{equation} \label{GrindEQ__58_}
\frac{1}{r^{2} } \frac{\partial }{\partial r} r^{2} \frac{\partial }{\partial r} \left(-\frac{m}{r} \right)=4\pi a^{3} m\delta (r).
\end{equation}
Therefore in order to extract the particle-like singular part of the solution further on it is convenient to introduce a new field function $\psi (r,\eta )$ such as [9]

\begin{equation} \label{GrindEQ__59_}
\delta \nu =-\delta \lambda =2\frac{\psi (r,\eta )-m(\eta )}{ar} \equiv 2\frac{\Phi (r,\eta )}{ar} ,                                                             (59)
\end{equation}
At that according to \eqref{GrindEQ__56_} the relation must be fulfilled

\begin{equation} \label{GrindEQ__60_}
\left|\mathop{\lim }\limits_{r\to 0} \frac{\psi }{r} \right|<\infty .
\end{equation}
Extracting the singular part of energy density in the right part of  Eq. \eqref{GrindEQ__43_}, substituting the function  $\delta \nu $ in the form \eqref{GrindEQ__59_} into Eqs. \eqref{GrindEQ__42_}-\eqref{GrindEQ__44_} and excluding the singular part subject to the relations \eqref{GrindEQ__58_} and \eqref{GrindEQ__60_} we get the linear equations set relatively the function $\Phi $ and perturbations of energy density and velocity

\begin{equation} \label{GrindEQ__61_}
\ddot{\Phi }+\frac{\dot{a}}{a} \dot{\Phi }-\frac{3}{2} (1+\kappa )\frac{\mathop{\dot{a}}\nolimits^{2} }{a^{2} } \Phi =4\pi ra^{3} \kappa \delta \varepsilon ,
\end{equation}

\begin{equation} \label{GrindEQ__62_}
3\frac{\dot{a}}{a} \dot{\Phi }-\psi ''=-4\pi ra^{3} \delta \varepsilon ,
\end{equation}

\begin{equation} \label{GrindEQ__63_}
\frac{\partial }{\partial r} \frac{\dot{\Phi }}{r} =-4\pi ra^{3} (1+\kappa )\varepsilon _{0} v.
\end{equation}
Multiplying Eq. \eqref{GrindEQ__62_} by $\kappa $ and adding its both parts to the corresponding parts of  Eq. \eqref{GrindEQ__61_} we obtain a closed equation

\begin{equation} \label{GrindEQ__64_}
\ddot{\Phi }+\frac{\dot{a}}{a} (1+3\kappa )\dot{\Phi }-\frac{3}{2} (1+\kappa )\frac{\mathop{\dot{a}}\nolimits^{2} }{a^{2} } \Phi -\kappa \psi ''=0.
\end{equation}
Then assuming according to \eqref{GrindEQ__59_}

\begin{equation} \label{GrindEQ__65_}
\Phi (r,\eta )=\psi (r,\eta )-m(\eta )
\end{equation}
And dividing the variables in Eq. \eqref{GrindEQ__64_}, we get two equations for the functions $m(\eta )$ and  $\psi (r,\eta )$

\begin{equation} \label{GrindEQ__66_}
\ddot{m}+\frac{\dot{a}}{a} (1+3\kappa )\dot{m}-\frac{3}{2} (1+\kappa )\frac{\mathop{\dot{a}}\nolimits^{2} }{a^{2} } m=\Theta (\eta );
\end{equation}

\begin{equation} \label{GrindEQ__67_}
\ddot{\psi }+\frac{\dot{a}}{a} (1+3\kappa )\dot{\psi }-\frac{3}{2} (1+\kappa )\frac{\mathop{\dot{a}}\nolimits^{2} }{a^{2} } \psi -\kappa \psi ''=-\Theta (\eta ),
\end{equation}
where $\Theta (\eta )$ is an arbitrary function of its argument.

\subsection{Basic theorem }

\noindent Further let $m=M(\Theta ,\eta )$ be the private Eq. \eqref{GrindEQ__66_} solution corresponding to the given function $\Theta (\eta )$. Then in consequence of Eq. \eqref{GrindEQ__67_} linearity  the function $\psi _{1} =-M(\Theta ,\eta )$.is a private solution of this equation.   Then Eq. \eqref{GrindEQ__67_}  general solution  can be written in the form

\begin{equation} \label{GrindEQ__68_}
\psi (r,\eta )=\psi _{0} (r,\eta )-M(\Theta ,\eta ),
\end{equation}
where  $\Psi (r,\eta )$ is general solution of the corresponding homogeneous equation

\begin{equation} \label{GrindEQ__69_}
\ddot{\Psi }+\frac{\dot{a}}{a} (1+3\kappa )\dot{\Psi }-\frac{3}{2} (1+\kappa )\frac{\mathop{\dot{a}}\nolimits^{2} }{a^{2} } \Psi -\kappa \Psi ''=0.
\end{equation}
Further on in consequence of the Eq. \eqref{GrindEQ__66_} linearity its general solution

\noindent is  sum of the corresponding homogeneous equation  $\mu (\eta )$ and private solution of inhomogeneous one

\begin{equation} \label{GrindEQ__70_}
m(\eta )=\mu (\eta )+M(\Theta ,\eta ).
\end{equation}
But then

\noindent

\begin{equation} \label{GrindEQ__71_}
\Phi (r,\eta )=\psi (r,\eta )-m(\eta )=\Psi (r,\eta )-\mu (\eta ),
\end{equation}
where $\mu (\eta )$ is the homogeneous equation general solution

\begin{equation} \label{GrindEQ__72_}
\ddot{\mu }+\frac{\dot{a}}{a} (1+3\kappa )\dot{\mu }-\frac{3}{2} (1+\kappa )\frac{\mathop{\dot{a}}\nolimits^{2} }{a^{2} } \mu =0
\end{equation}
The rest equations of the set \eqref{GrindEQ__61_}-\eqref{GrindEQ__63_} describe the evolution of nonsingular part of energy density and perturbations velocity

\begin{equation} \label{GrindEQ__73_}
\delta \varepsilon =-\frac{1}{4\pi ra^{3} } \left(3\frac{\dot{a}}{a} (\dot{\Psi }-\dot{\mu })-\Psi ''\right),
\end{equation}

\begin{equation} \label{GrindEQ__74_}
\frac{\partial }{\partial r} \frac{\dot{\Psi }-\dot{\mu }}{r} =-4\pi ra^{3} (1+\kappa )\varepsilon _{0} v.
\end{equation}
Thus we proved the theorem

\noindent \textbf{Theorem}. \textit{Linear spherically symmetric perturbations of Friedmann metrics are described by a set of two independent linear homogeneous equations} \textit{\eqref{GrindEQ__69_} and \eqref{GrindEQ__72_} relatively two functions }$\mu (\eta )$\textit{ and }$\psi (r,\eta )$ \textit{which are nonsingular at the origin of the coordinates. Spherically symmetric perturbations of energy density and velocity are determined through metrics perturbations by the relations \eqref{GrindEQ__73_}-\eqref{GrindEQ__74_} }.

\noindent \textit{At $\kappa >0$ the homogeneous Eq.
\eqref{GrindEQ__69_} is hyperbolic, at $\kappa <0$ it is elliptic,
at $\kappa =0$ this equation coincides with Eq.
\eqref{GrindEQ__72_}. }

\section{Evolutional equations for
perturbations at constant barotrop coefficient }

\subsection{Mass evolution of particle-like source }

\noindent Let us study the cosmological mass evolution of particle-like source. Passing on from the variable $\eta $ to the variable $a(\eta )$ subject to the relation \eqref{GrindEQ__42_} in the mass evolution equation \eqref{GrindEQ__72_}  let us reduce it to the form

\begin{equation} \label{GrindEQ__75_}
\frac{d^{2} \mu }{da^{2} } +\frac{3}{2} \frac{1+\kappa }{a} \frac{d\mu }{da} -\frac{3}{2} (1+\kappa )\frac{\mu }{a^{2} } =0.
\end{equation}
The general solution of this equation can be easily obtained

\begin{equation} \label{GrindEQ__76_}
\mu =C_{+} a+C_{-} a^{-{\tfrac{3}{2}} (1+\kappa )} ,
\end{equation}
where $C_{+} $ and $C_{-} $ are arbitrary constants. The $C_{+} $ coefficient term in this solution corresponds to accretion processes, the $C_{-} $ coefficient one corresponds to evaporation processes. Substituting in  Eq.\eqref{GrindEQ__76_} the scale factor relation to the time variable $\eta $ \eqref{GrindEQ__32_} and using the relation \eqref{GrindEQ__34_} between the time variable $\eta $ and the physical time  $t$ we obtain in an explicit form the law of mass evolution of particle-like source

\begin{equation} \label{GrindEQ__77_}
\mu =\mathop{\tilde{C}}\nolimits_{+} t^{{\tfrac{2}{3(1+\kappa )}} } +\mathop{\tilde{C}}\nolimits_{-} t^{-1} ,\quad (1+\kappa )\ne 0,
\end{equation}
Where $\mathop{\tilde{C}}\nolimits_{+} $ and  $\mathop{\tilde{C}}\nolimits_{-} $ are new arbitrary constants. From the solutions \eqref{GrindEQ__76_} and \eqref{GrindEQ__77_} it is seen $C_{-} =\mathop{\tilde{C}}\nolimits_{-} 0$ corresponds to the final particle mass at the moment of time $t=0$. The solutions \eqref{GrindEQ__76_} - \eqref{GrindEQ__77_} summarize the solutions obtained in the previous papers [3]-[4] for the two private values of adiabatic curve coefficient $\kappa =0$ and $\kappa =1/3$which correspond to nonrelativistic and  ultrarelativistic equation of state correspondingly. In the pointed out cases we get from \eqref{GrindEQ__77_}

\begin{equation} \label{GrindEQ__78_}
\mu =\mathop{\tilde{C}}\nolimits_{+} t^{{\tfrac{2}{3}} } +\mathop{\tilde{C}}\nolimits_{-} t^{-1} ,\quad \kappa =0,
\end{equation}

\begin{equation} \label{GrindEQ__79_}
\mu =\mathop{\tilde{C}}\nolimits_{+} t^{{\tfrac{1}{2}} } +\mathop{\tilde{C}}\nolimits_{-} t^{-1} ,\quad \kappa =\frac{1}{3} ,
\end{equation}
Let us study a numerical example. Let the particle-like source mass equal to Planck mass at the moment of time $t=t_{Pl} $ then at the contemporary moment of time  $t\sim 10^{60} t_{Pl} $ according to \eqref{GrindEQ__77_} the contemporary mass of ``particle'' varies within the limits $10^{-18} M_{\odot } $ at $\kappa =1$ till $10^{2} M_{\odot } $ at $\kappa =0$. At  negative values of the adiabate curve index the perturbation mass rapidly increases and at $\kappa \approx -0,5$ in order of value it is compared with  the visible universe mass (see Fig.1).

\noindent

\noindent In the case of the inflation solution \eqref{GrindEQ__36_} the evolutional equation \eqref{GrindEQ__72_} takes the form

\begin{equation} \label{GrindEQ__80_}
\ddot{\mu }=-\frac{2}{\eta } \dot{\mu }=0.
\end{equation}
Solving \eqref{GrindEQ__80_} we obtain

\begin{equation} \label{GrindEQ__81_}
m=C_{+} +\frac{C_{-} }{\eta } .
\end{equation}
Now using the relation of the time constant $\eta $ to the physical time $t$ for the case of inflation  \eqref{GrindEQ__36_} for this case finally we get

\begin{equation} \label{GrindEQ__82_}
\mu =C_{+} +C_{-} e^{-\Lambda t} .
\end{equation}
In this case $C_{-} =0$ corresponds to the final mass at $t\to -\infty $again.

\noindent \textbf{\includegraphics[bb=0mm 0mm 208mm 296mm, width=99.3mm, height=91.5mm, viewport=3mm 4mm 205mm 292mm]{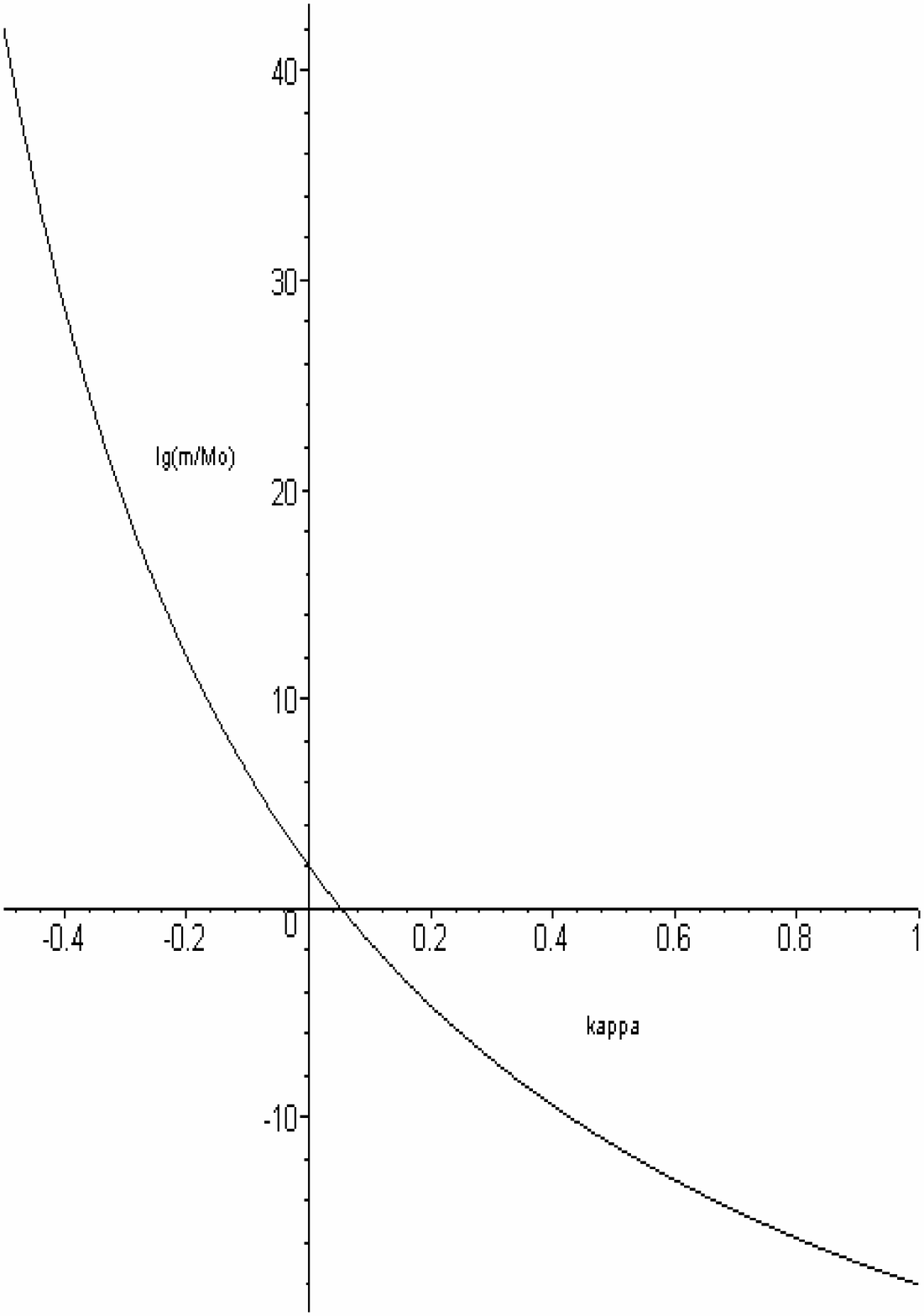}}

\noindent \textbf{Fig.1.} Relationship of contemporary mass value
of particle-like source in Friedmann world which had Planck mass
at Planck moment of time to barotrop coefficient $\kappa $. On the
ordinate axis the mass logarithm values are marked off in Solar
mass units $M_{\odot } \approx 2\cdot 10^{33} $ \textit{g.}

\subsection{Evolutional Equation for nonsingular mode of
perturbations }

\noindent From the general nonperturbed solution \eqref{GrindEQ__32_} we obtain useful ratios

\begin{equation} \label{GrindEQ__83_}
\frac{\dot{a}}{a} =\frac{2}{3\kappa +1} \, \frac{1}{\eta } ;\quad \frac{\ddot{a}}{a} =\frac{2(1-3\kappa )}{(3\kappa +1)^{2} } \, \frac{1}{\eta ^{2} } ,\quad (1+\kappa \ne 0),
\end{equation}
using which we reduce Eq. \eqref{GrindEQ__69_} for  nonsingular mode of perturbations to the form

\begin{equation} \label{GrindEQ__84_}
\ddot{\Psi }+\frac{2}{\eta } \dot{\Psi }-\frac{6(1+\kappa )}{(1+3\kappa )^{2} } \frac{\Psi }{\eta ^{2} } -\kappa \Psi ''=0.
\end{equation}

\noindent In the case of $(1+\kappa )=0$ Eq. \eqref{GrindEQ__74_} is no longer an equation for defining a radial velocity of perturbations, but becomes a differential equation relatively the function  $\psi $

\begin{equation} \label{GrindEQ__85_}
\frac{\partial }{\partial r} \frac{\dot{\Psi }-\dot{\mu }}{r} =0.
\end{equation}
It is necessary to solve this equation simultaneously with Eq. \eqref{GrindEQ__84_} which in this case takes the form

\begin{equation} \label{GrindEQ__86_}
\ddot{\Psi }+\frac{2}{\eta } \dot{\Psi }+\Psi ''=0.
\end{equation}
\subsection{General solution of evolution equation for
nonsingular mode of perturbations }

\noindent Assuming

\[\Psi (r,\eta )=R(r)\Theta (\eta )\]
and dividing variables in Eq. \eqref{GrindEQ__84_} we obtain ordinary differential equations

\begin{equation} \label{GrindEQ__87_}
\kappa R''+e\alpha ^{2} R=0,
\end{equation}

\begin{equation} \label{GrindEQ__88_}
\eta ^{2} \ddot{\Theta }+2\eta \dot{\Theta }+\left[e\alpha ^{2} \eta ^{2} -6\frac{1+\kappa }{(1+3\kappa )^{2} } \right]\Theta =0.
\end{equation}
To guarantee similarity of the solutions the sign of the division constant should be opposite to the barotrop coefficient

\begin{equation} \label{GrindEQ__89_}
e=-{\rm sgn}(\kappa ).
\end{equation}
Solving Eq. \eqref{GrindEQ__87_} we get

\begin{equation} \label{GrindEQ__90_}
R=C_{1} \sin \frac{\alpha }{\sqrt{\kappa } } r+C_{2} \cos \frac{\alpha }{\sqrt{\kappa } } r.
\end{equation}
In order the function do not contain peculiarities at the origin of the coordinates $\Psi (r,\eta )$ (condition (?)) it is necessary and sufficient in the solution \eqref{GrindEQ__90_} $C_{2} =0$, thus

\begin{equation} \label{GrindEQ__91_}
R(r)=C(\alpha )\sin \frac{\alpha }{\sqrt{|\kappa |} } r,
\end{equation}
where  $C(\alpha )$is an arbitrary constant.

\noindent At  $\kappa >0$  Eq. \eqref{GrindEQ__88_} has its solution

\begin{equation} \label{GrindEQ__92_}
\Theta (\eta )=\frac{\mathop{\tilde{C}}\nolimits_{1} }{\sqrt{\eta } } J_{s} (\alpha \eta )+\frac{\mathop{\tilde{C}}\nolimits_{2} }{\sqrt{\eta } } Y_{s} (\alpha \eta ),
\end{equation}
where $J_{s} (z)$ and $Y_{s} (z)$ are the Bessel functions of the first and second genuses respectively and

\begin{equation} \label{GrindEQ__93_}
s=\frac{1}{2} \left|\frac{5+3\kappa }{1+3\kappa } \right|\frac{1}{2} ,\quad \kappa \in [-1,+\infty ).
\end{equation}
At $\kappa <0$ the Eq. \eqref{GrindEQ__88_} solution is expressed by the way of the Bessel functions of imaginary argument $I_{s} (z)$ and $K_{s} (z)$  (see, e.g. [10]):

\begin{equation} \label{GrindEQ__94_}
\Theta (\eta )=\frac{\mathop{\tilde{C}}\nolimits_{1} }{\sqrt{\eta } } I_{s} (\alpha \eta )+\frac{\mathop{\tilde{C}}\nolimits_{2} }{\sqrt{\eta } } K_{s} (\alpha \eta ).
\end{equation}
As far as the functions $J_{s} (z)/\sqrt{z} $ and  $K_{s} (z)/\sqrt{z} $ tend to infinity at $z\to 0$ in order to get restricted ones at $\eta \to 0$ it is necessary to set in the formulae \eqref{GrindEQ__92_} and \eqref{GrindEQ__94_}

\begin{equation} \label{GrindEQ__95_}
\mathop{\tilde{C}}\nolimits_{2} =0.
\end{equation}
Thus nonsingular solution of evolutionary equation for perturbations can be written in the form

\begin{equation} \label{GrindEQ__96_}
\Phi (r,\eta )=\frac{1}{\sqrt{\eta } } \int _{0}^{\infty } \, C(\alpha )\sin \frac{\alpha }{\sqrt{|\kappa |} } rJ_{s} (\alpha \eta )d\alpha ,\quad \kappa >0.
\end{equation}
At $\kappa <0$ we obtain simultaneously

\begin{equation} \label{GrindEQ__97_}
\Phi (r,\eta )=\frac{1}{\sqrt{\eta } } \int _{0}^{\infty } \, C(\alpha )\sin \frac{\alpha }{\sqrt{|\kappa |} } rI_{s} (\alpha \eta )d\alpha ,\quad \kappa <0.
\end{equation}

\end{document}